\documentclass[preprint2]{aastex}

\newcommand{\etal}{et al.\ }

\slugcomment{Submitted to ApJ}
\shorttitle{X-ray-brightest T-Tauri stars in $\rho$ Oph}
\shortauthors{Imanishi, Tsujimoto, \& Koyama}

\received{2001 December 17}
\begin{document}

\title{{\it Chandra} and {\it ASCA} Observations of the X-ray-brightest
T-Tauri Stars in the $\rho$ Ophiuchi Cloud}

\author{Kensuke~Imanishi, Masahiro~Tsujimoto, and Katsuji~Koyama}
\affil{Department of Physics, Graduate School of Science, Kyoto
University, Sakyo-ku, Kyoto, 606-8502, Japan}
\email{kensuke@cr.scphys.kyoto-u.ac.jp,
tsujimot@cr.scphys.kyoto-u.ac.jp, koyama@cr.scphys.kyoto-u.ac.jp}

\begin{abstract}

We present the {\it Chandra} ACIS and {\it ASCA} GIS results for a
series of four long-term observations on DoAr 21, ROXs 21 and ROXs 31;
the X-ray brightest T-Tauri stars (TTSs) in the $\rho$ Ophiuchi
cloud. In the four observations with a net exposure of $\sim$ 600 ksec,
we found six, three and two flares from DoAr 21, ROXs 21 and ROXs 31,
respectively; hence the flare rate is fairly high. The spectra of DoAr
21 are well fitted with a single-temperature plasma model, while those
of ROXs 21 and ROXs 31 need an additional soft plasma component. Since
DoAr 21 is younger ($\sim$10$^5$ yr) than ROXs 21 and ROXs 31
($\sim$10$^6$ yr), these results may indicate that the soft component
gradually increases as T-Tauri stars age.  The abundances are generally
sub-solar and vary from element to element. Both high-FIP (first
ionization potential) and low-FIP elements show enhancement over the
mean abundances.  An unusual giant flare is detected from ROXs 31. The
peak luminosity and temperature are $\sim$10$^{33}$ ergs s$^{-1}$ and
$\sim$10 keV, respectively. The temperature reaches its peak value
before the flux maximum, and is nearly constant (4--5 keV) during the
decay phase, indicating successive energy release during the flare. The
abundances and absorption show dramatic variability from the quiescent
to flare phase.

\end{abstract}

%\keywords{stars: abundances --- stars: coronae --- stars: individual
%(DoAr 21, ROXs 21, ROXs 31) --- stars: pre-main sequence --- X-rays:
%stars}

\section{INTRODUCTION}

Low-mass stars in the pre-main sequence (PMS) stage are classified into
four types based on their infrared (IR) to sub-mm spectral energy
distributions (SEDs); molecular cloud cores through protostars are
represented by class 0 and I SEDs, while classical and weak-line T-Tauri
stars (CTTSs and WTTSs) exhibit class II and III SEDs
\citep{Shu1987,Andre1994}.

The {\it Einstein} satellite first revealed that T-Tauri stars (TTSs =
CTTSs and WTTSs) emit intense and time variable X-rays with occasional
rapid flares \citep{Feigelson1981,Montmerle1983}. These properties are
consistent with a scenario of enhanced solar-type activity, attributable
to magnetic dynamo processes. Successive X-ray satellites such as {\it
ROSAT}, {\it ASCA}, and {\it Chandra} have detected X-ray emission from
many PMS stars, even from the youngest class 0 stage \citep{Tsuboi2001},
and suggest that X-ray emission from low-mass PMS stars is caused by
magnetic activity. These results, however, rely mainly on short duration
observations; neither long-term nor detailed behavior of the X-ray
spectrum and timing has been well studied. This paper hence addresses
three topics on TTSs; (1) the duty ratio of the X-ray flare, (2) the
long-term plasma structure and variability, and (3) the chemical
compositions in the quiescent and flare phases.

(1) \citet{Stelzer2000} compiled {\it ROSAT} PSPC observations of the
Taurus-Auriga-Perseus region and found that the flare rate may depend on
the stellar age, rotation period, and/or multiplicity; stars with
younger ages, shorter rotation periods, and located in binary systems
tend to show more frequent flares. Hence, the flare rate may help to
reveal the physical conditions of PMS stars.

(2) \citet{Preibisch1997} fitted the {\it ROSAT} PSPC spectra of
$\sim$60 PMS stars with several thermal plasma models and found that
multi-temperature components are needed for most of the
spectra. \citet{Tsujimoto2001} also reported that about two-thirds of
the {\it Chandra} spectra of TTSs in the Orion molecular cloud tend to
be well fitted by multi-temperature plasma models as compared with
single temperature plasma models for younger objects (protostars). The
multi-temperature structure may also depend on the activity phase of
each star; higher temperature plasma becomes dominant in a flare
compared with a quiescent phase. Thus long-term monitoring of both flare
and quiescent phases is crucial for the study of the origin and
structure of the multi-temperature plasma.

(3) {\it ASCA} found that the mean chemical composition of
X-ray-emitting plasma in young stellar objects (YSOs) is generally
sub-solar and varies from star to star
\citep{Kamata1997,Ozawa2000}. These results may be misleading, however,
because the mean-abundance determination is biased toward elements with
strong emission lines in the relevant energy range and largely depends
on the plasma temperature. Moreover, the abundances of the stellar
corona may be different from those in the photosphere, depending on the
first ionization potential (FIP) of the relevant elements. For instance,
in a solar corona, elements with low FIP are overabundant relative to
the solar photospheric values (``the FIP effect'', Feldman 1992). In
some of the stellar coronae, however, the abundances of high-FIP
elements are larger than the others (the Inverse FIP = IFIP effect). The
IFIP effect appears in X-ray active stars, while the FIP effect is
generally seen in less active stars \citep{Audard2001, Brinkman2001,
Guedel2001a, Guedel2001b, Guedel2001c, Huenemoerder2001,
Kastner2001}. The dependence of abundance on the X-ray activity is also
found within individual stars; the enhancements in abundance are seen
from quiescent to flare phases, or even within a single flare phase
(e.g. HR1099, Audard et al.\ 2001; V773 Tau, Tsuboi et al.\ 1998). Due
to the limited statistics and samples, however, it is unclear whether or
not abundance variations between quiescent and flare phases are common,
and whether or not FIP/IFIP effects are present in low-mass YSOs.

To investigate the above-mentioned subjects, we selected the three X-ray
brightest TTSs; DoAr 21 (ROXs 8), ROXs 21 and ROXs 31 in the $\rho$
Ophiuchi cloud ($\rho$ Oph) located at a distance of 145 pc
\citep{deZeeuw1999}, because all three have publicly available long-term
observations with {\it Chandra} and {\it ASCA}. These TTSs were first
recognized as bright X-ray sources with {\it Einstein}
\citep{Montmerle1983}, and were identified with optical K--M stars
having weak H$\alpha$ emission \citep{Bouvier1992}, hence classified as
WTTS (class III). \citet{Simon1995} observed the lunar occultation of
these WTTSs at infrared wavelengths and found that ROXs 21 and ROXs 31
are binary systems with separation angles of 0\farcs3 and 0\farcs48,
respectively. The rotational period of ROXs 21 is reported to be 1.39
days \citep{Shevchenko1998}.  Since no continuum emission is found at
1.3 mm \citep{Andre1994}, the circumstellar envelope of these TTSs is
thought to have already disappeared, while DoAr 21 may still have an
accretion disk, as suggested by the detection of near-infrared (NIR)
polarized emission \citep{Ageorges1997}.  Using the theoretical
evolutionary tracks in the H-R diagram, the age of DoAr 21 is estimated
to be $\sim$10$^5$ yr, which is younger than that of ROXs 21 and ROXs 31
($\sim$10$^6$ yr) \citep{Nurnberger1998}. Finally, DoAr 21 and ROXs 31
show variable radio emission, which is probably due to the
gyro-synchrotron emission induced by the surface magnetic field
\citep{Stine1988}.

Early results with the {\it Chandra} and {\it ASCA} observations
\citep{Koyama1994,Kamata1997,Tsuboi1999,Tsuboi2000,Skinner2000,Imanishi2001}
revealed that these particular TTSs have luminosities of
$\gtrsim$10$^{30}$ ergs s$^{-1}$ and relatively low absorption ($N_{\rm
H}$ $\lesssim$ 10$^{22}$ cm$^{-2}$). Thus they can provide enough counts
to measure their temporal and spectral properties in great detail.

\section{OBSERVATIONS AND DATA REDUCTION}

The {\it Chandra} \citep{Weisskopf1996} observation (hereafter, obs C1)
of these TTSs was made with five front-illuminated X-ray CCDs, ACIS-I0,
I1, I2, I3, and S2, in the 0.2--10 keV band. The level 2 data were
retrieved from the {\it Chandra} X-ray Center (CXC) archive. X-ray
events are selected using the {\it ASCA} grades 0, 2, 3, 4, and 6. An
effective exposure time of $\approx$100 ks is obtained (Table
\ref{tab:obs}). Another long exposure observation of the $\rho$ Oph core
A, which contains DoAr 21 in the ACIS-I3, is also in the archive (obs
C2), but this data suffer from the photon pile-up and therefore were not
used.

Three observations (obs A1, A2, and A3) were carried out with the two
Gas Imaging Spectrometers (GISs) \citep{Ohashi1996} and the two
Solid-state Imaging Spectrometers (SISs) \citep{Burke1991} onboard {\it
ASCA} \citep{Tanaka1994}, at the foci of the X-ray telescopes (XRTs)
\citep{Serlemitsos1995} sensitive to photons in 0.4--10 keV. However,
the TTS sources were located outside of, or at the edge of the SIS's
field of view in the majority of the observations; therefore we do not
use the SIS data. We retrieved the {\it ASCA} unscreened data from the
HEASARC Online Service. Data taken at a geomagnetic cutoff rigidity
lower than 4 GV, at an elevation angle less than 5$^\circ$ from the
Earth, and during passage through the South Atlantic Anomaly were
filtered out. Particle events were also removed using the rise-time
discrimination method. The total available exposure times were
$\approx$38 ks (obs A1), 93 ks (A2), and 75 ks (A3) (Table
\ref{tab:obs}).

\section{ANALYSIS AND RESULTS}

\subsection{Extraction of Source and Background Photons}

The positions of NIR counterparts \citep{Barsony1997} were used to
fine-tune the {\it Chandra} coordinates. We then extracted the source
photons using an ellipse with major and minor axes of 60\farcs0 and
35\farcs0, respectively, for DoAr 21, and from respective circles of
11\farcs7 and 12\farcs0 radius for ROXs 21 and ROXs31, with the regions
centered at the NIR positions \citep{Barsony1997}. The relatively large
source areas are due to large off-axis point spread functions (PSFs);
DoAr 21 lies in ACIS-S2 $\sim$20$'$ away from the optical axis, while
ROXs 21 and ROXs 31 are at the edge of ACIS-I1 with off-axis angles of
$\sim$7--10$'$. The large off-axis angle (and hence large PSF) for the
brightest source DoAr21 mitigates photon pile-up, which in fact is a
problem for the obs C2 case. The background region for DoAr 21 was taken
from a source-free 19 arcmin$^2$ region on ACIS-S2, while that for ROXs
21 and ROXs 31 was taken from a 63 arcmin$^2$ region on the ACIS-I
array. For the ASCA GIS data (obs A1--A3) the source photons were
extracted from 3$'$ radii circles. The background regions were also
taken using 3$'$ radii circles from nearby source-free regions.

\subsection{Time Variability}

\begin{figure}
 \figurenum{1}
 \epsscale{0.32}
\plotone{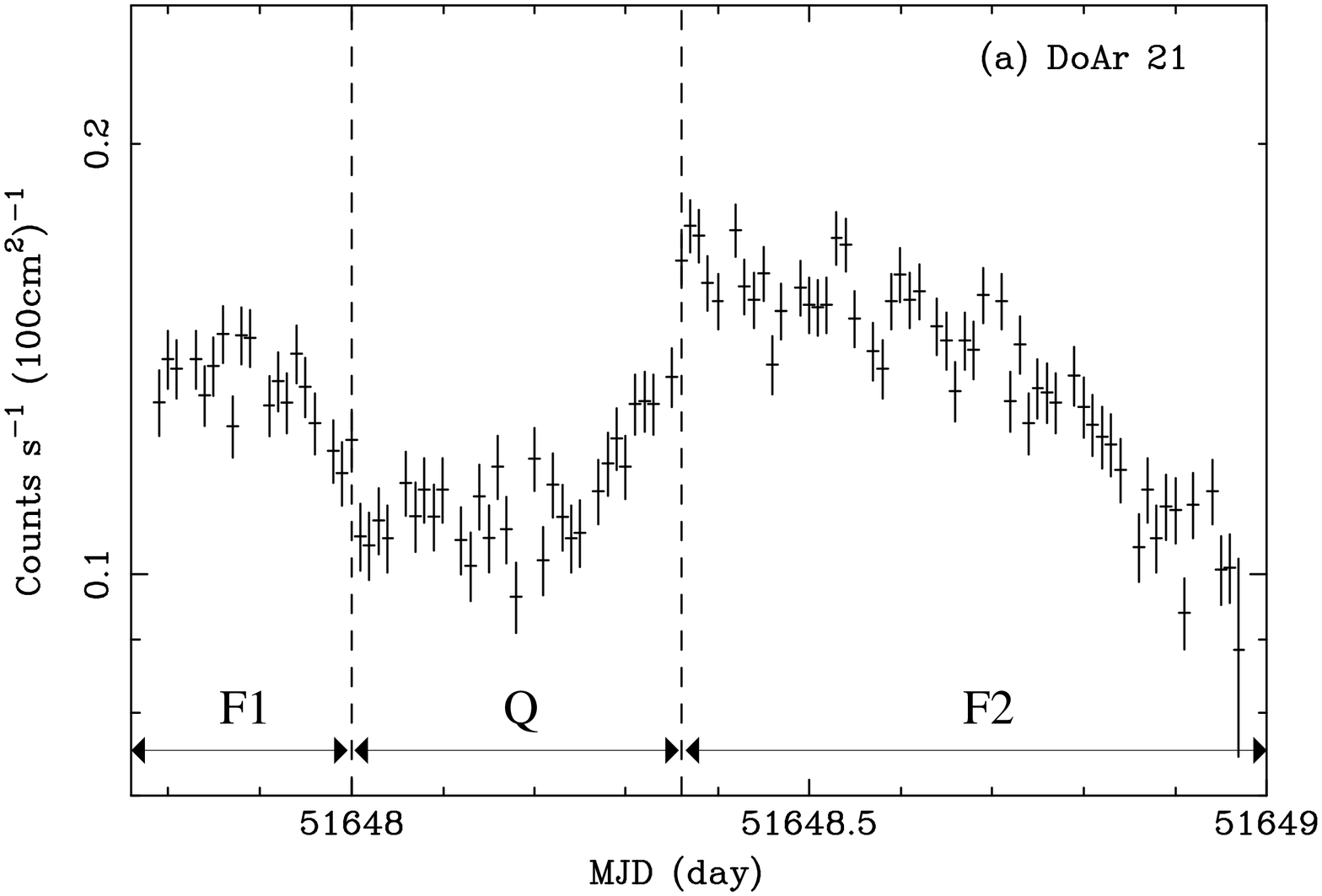}
\plotone{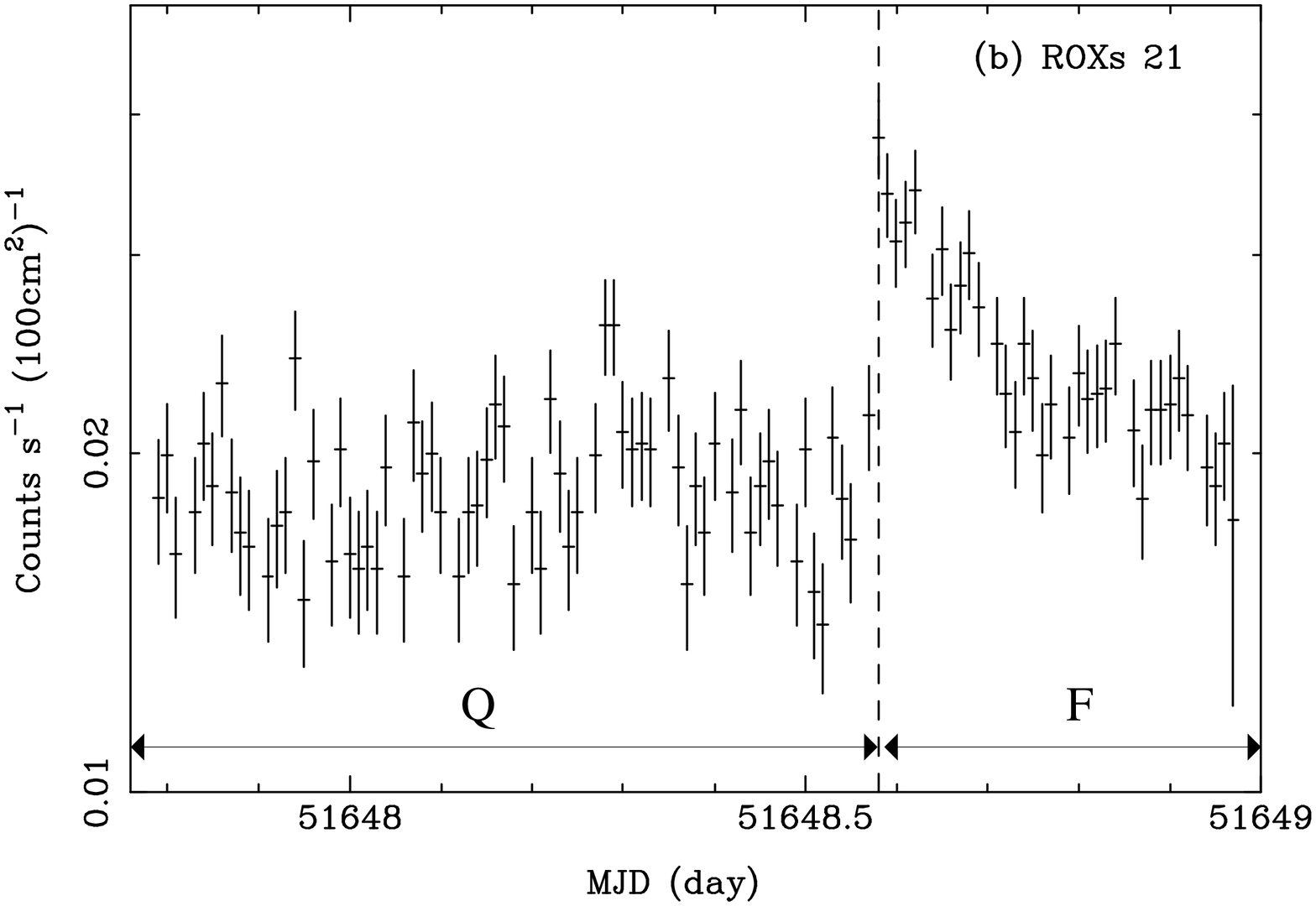}
\plotone{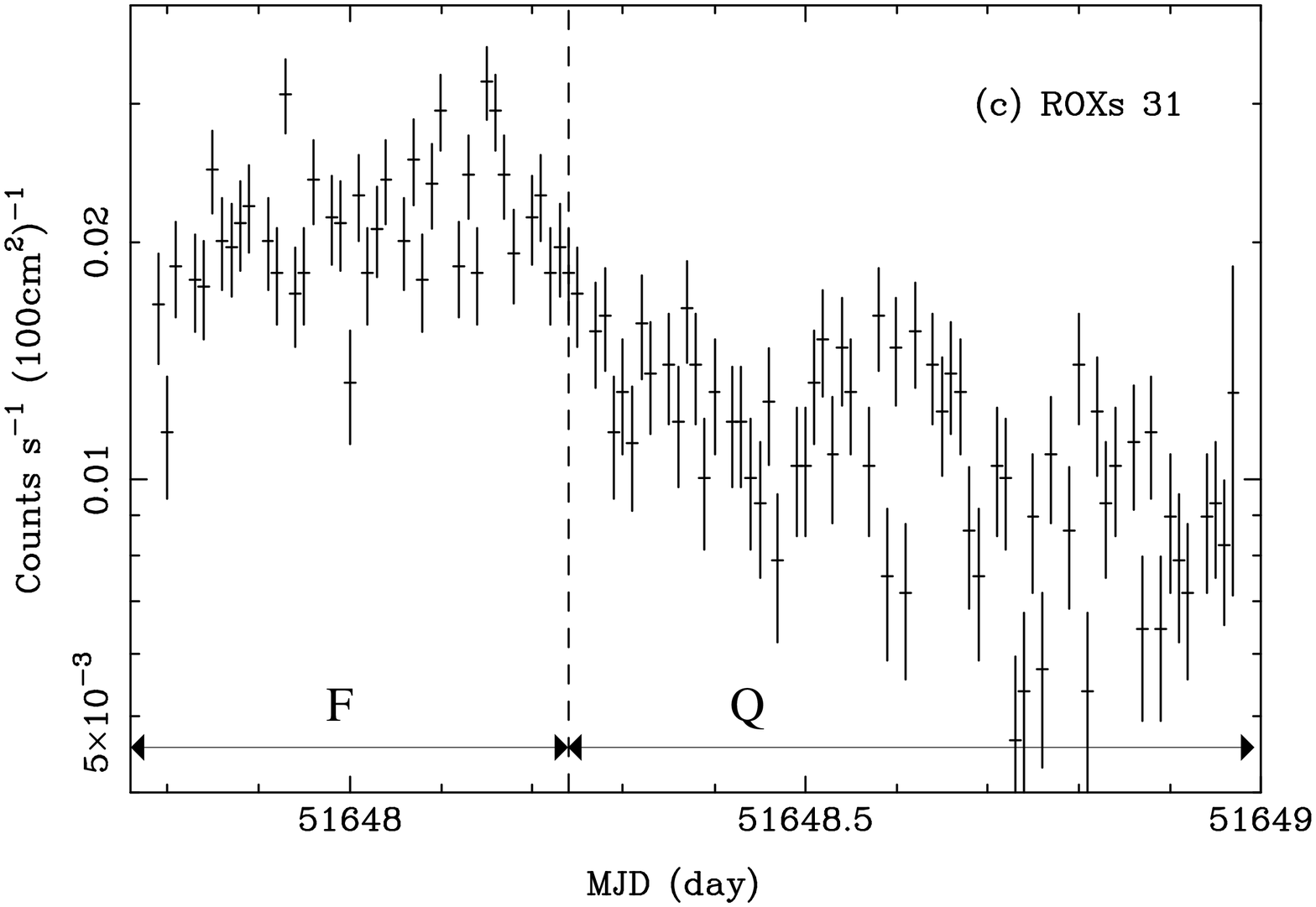}
 %
 % KI: fixed the word; horizontal -> vertical
 %
 \caption[f1a.eps,f1b.eps,f1c.eps]{Light curves of (a) DoAr 21, (b) ROXs
 21, and (c) ROXs 31 in 0.5--9.0 keV obtained with {\it Chandra} ACIS
 (obs C1). The vertical axis is the X-ray count rate normalized in the
 effective area of 100 cm$^2$ at 1 keV. The lower arrows indicate the
 times of the flare and quiescent phases (see Table
 \ref{tab:spec}). \label{fig:lc_c1}}
\end{figure}

\begin{figure}
 \figurenum{2}
 \epsscale{0.32}
\plotone{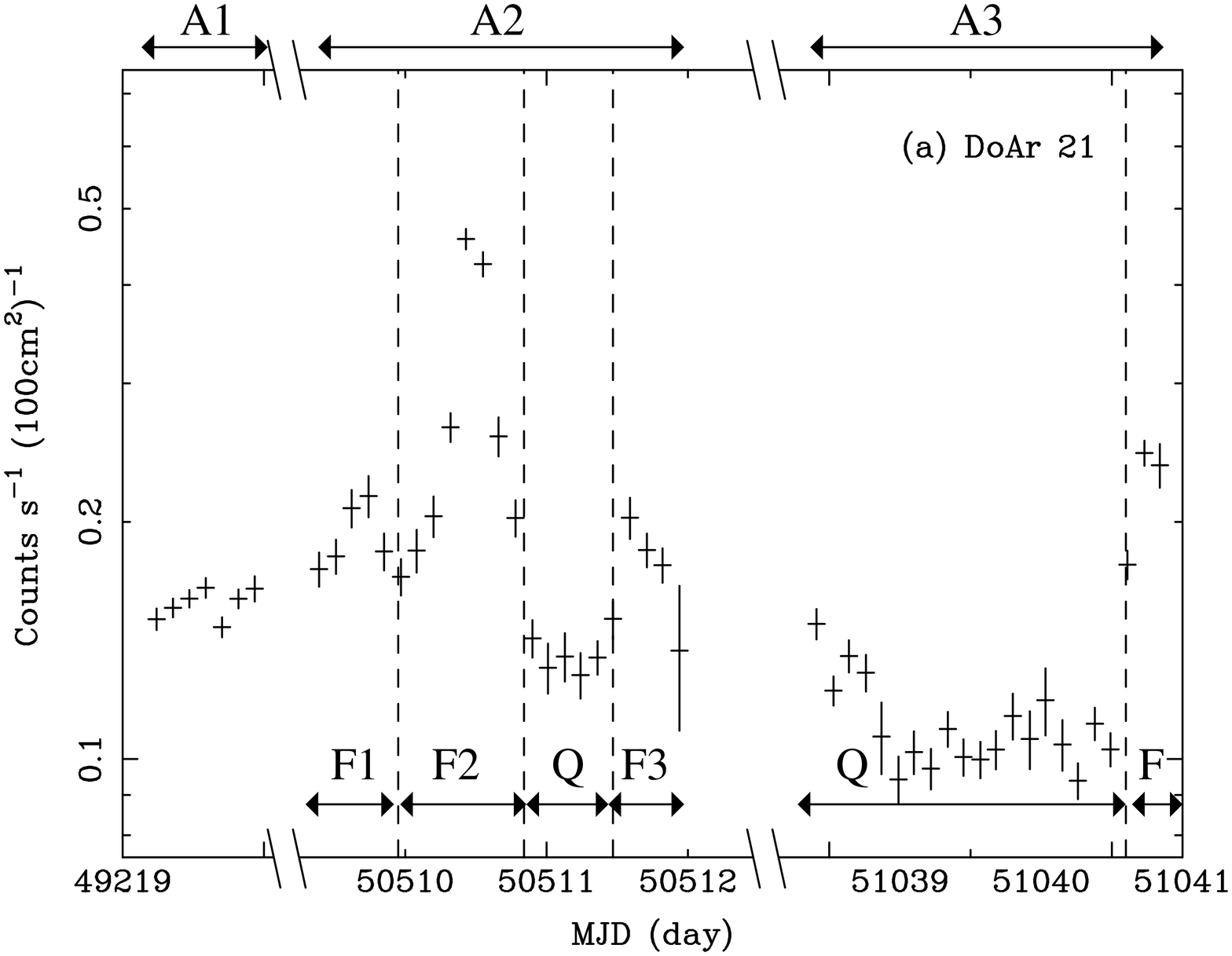}
\plotone{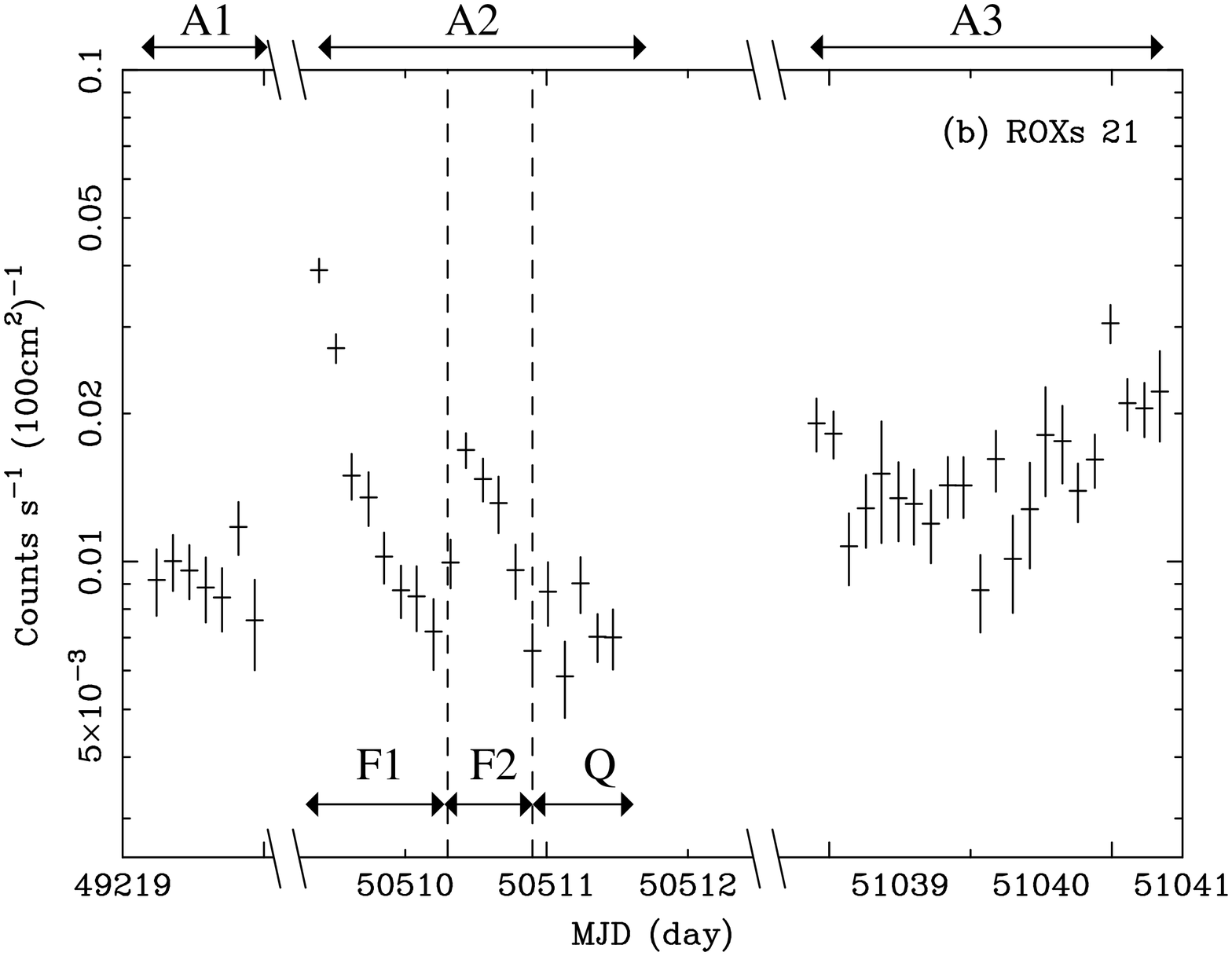}
\plotone{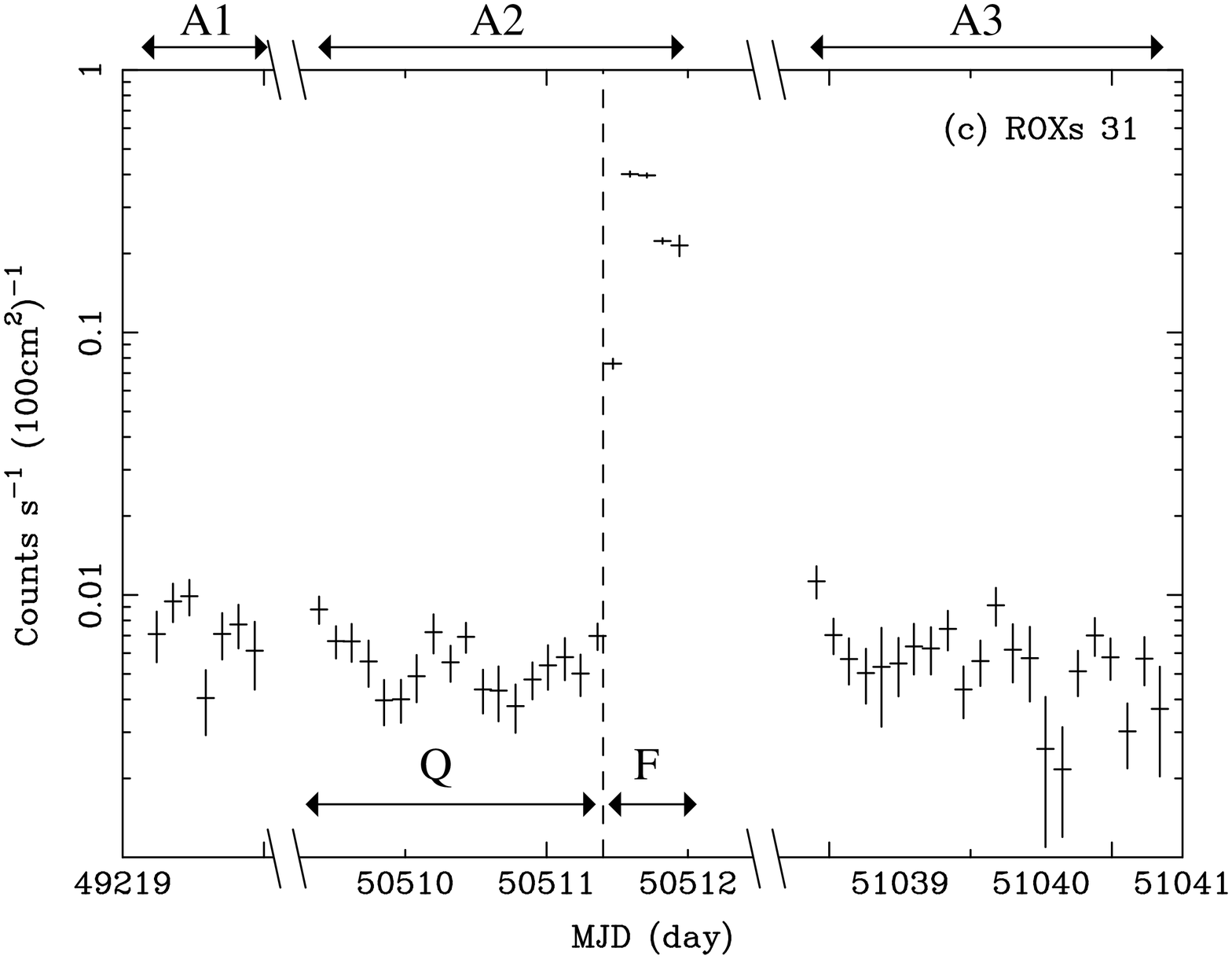}
 %
 % KI: fixed the word; horizontal -> vertical
 %
 \caption[f2a.eps,f2b.eps,f2c.eps]{Light curves of (a) DoAr 21 in
 0.5--9.0 keV, and of (b) ROXs 21 and (c) ROXs 31 in 0.5--1.5 keV,
 obtained with {\it ASCA} GIS (obs A1--A3). The vertical axis is the
 X-ray count rate normalized in the effective area of 100 cm$^2$ at 1
 keV. The lower arrows indicate the times of the flare and quiescent
 phases. \label{fig:lc_a1-3}}
\end{figure}

Figure \ref{fig:lc_c1} and \ref{fig:lc_a1-3} show the X-ray light curves
(no background subtraction) of (a) DoAr 21, (b) ROXs 21, and (c) ROXs 31
obtained with the {\it Chandra} ACIS (obs C1) and the {\it ASCA} GIS
(obs A1--A3), respectively. The photon counts are normalized by the
effective area at 1 keV, for comparison with the different instruments
and observation epochs. The time bins of each data point are 10$^3$ s
and 10$^4$ s for ACIS (Figure \ref{fig:lc_c1}) and GIS (Figure
\ref{fig:lc_a1-3}), respectively. The energy band of the ACIS and GIS
data for DoAr 21 are taken to be 0.5--9.0 keV. However, the GIS light
curves for ROXs 21 and ROXs 31 are limited in the soft energy band of
0.5--1.5 keV in order to avoid possible contamination from a nearby
variable hard X-ray source, YLW 15A
\citep{Tsuboi1999,Tsuboi2000,Imanishi2001}. To verify the energy band,
we extracted the ACIS-I spectrum of YLW 15A and fit it with an absorbed
thin thermal plasma model. We find a best-fitted absorption of $N_{\rm
H}$ = 4.4$\times$10$^{22}$ cm$^{-2}$ \citep{Imanishi2001}, which reduces
the X-ray emission below 1.5 keV to less than 3\% of the 1.5--9 keV
flux.  Hence possible contamination from YLW 15 A to the GIS source
circles of ROXs 21 and ROXs 31 is insignificant in the 0.5--1.5 keV
band. No data is shown in the ROXs 21 light curve (Figure
\ref{fig:lc_a1-3}b) from MJD = 50511.55--50512.20 due to significant
contamination from an exceptionally giant flare of ROXs 31 (see Figure
\ref{fig:lc_a1-3}c).

For a unified study, we conventionally define a flare using the
following criterion: at least two consecutive time bins must have larger
fluxes than the quiescent bin by $\ge 5\sigma$-level, where the
quiescent bin is 
%
%KI: change a phrase
%
the minimum time bin in each observation.  Under this criterion, we
detect six, three, and two flares from DoAr 21, ROXs 21 and ROXs 31,
respectively. These are labelled as ``F'', ``F1'', ``F2'', and ``F3''
with arrows in Figure \ref{fig:lc_c1} and \ref{fig:lc_a1-3}, while the
quiescent phases are indicated by ``Q''. An extremely large flare with
the maximum flux of $\sim$100 times higher than the quiescent level is
detected from ROXs 31 (Figure \ref{fig:lc_a1-3}c) and is separately
treated in \S3.4.

\subsection{X-ray Spectra}

\begin{figure}
 \figurenum{3}
 \epsscale{0.32}
\plotone{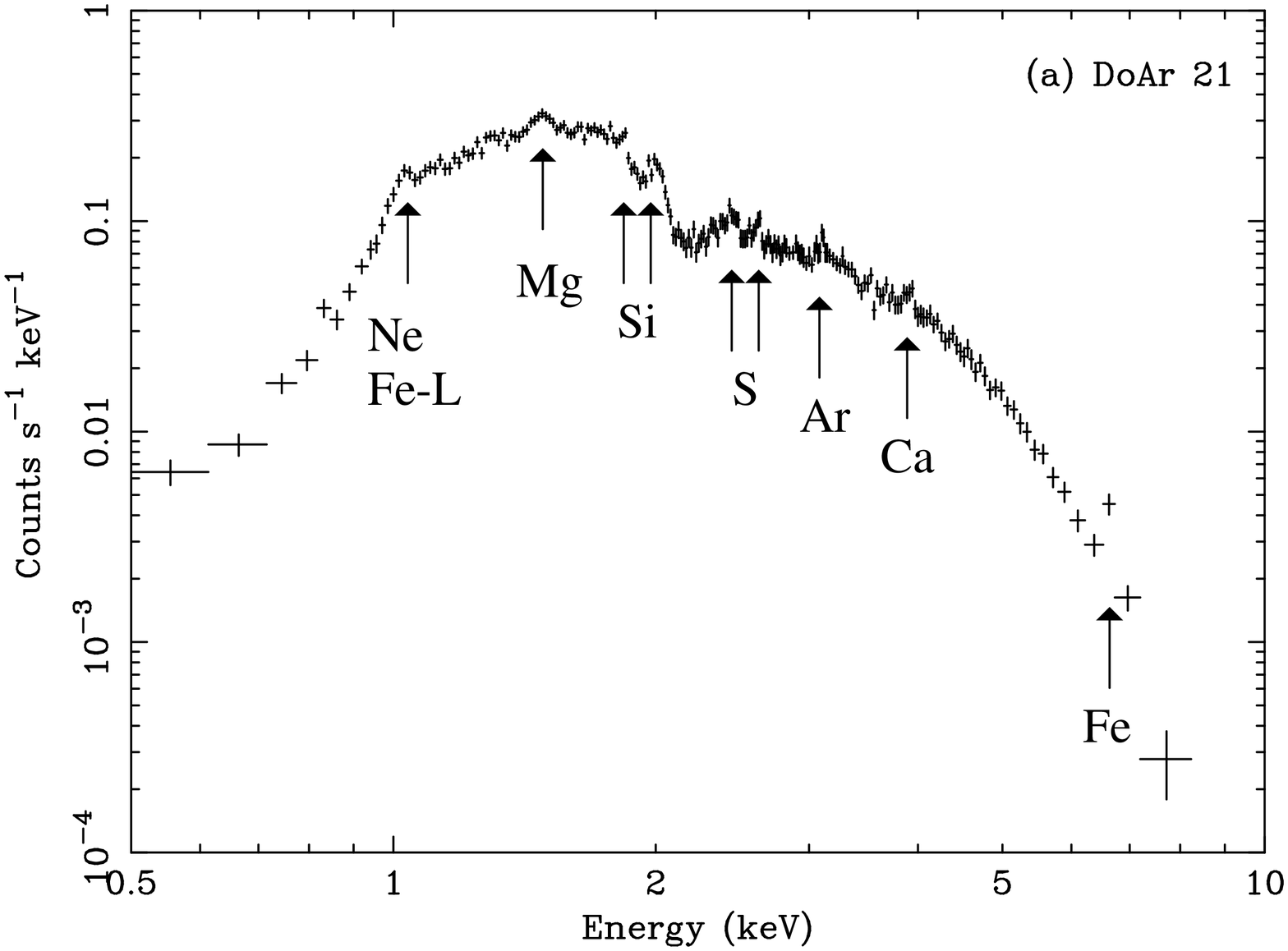}
\plotone{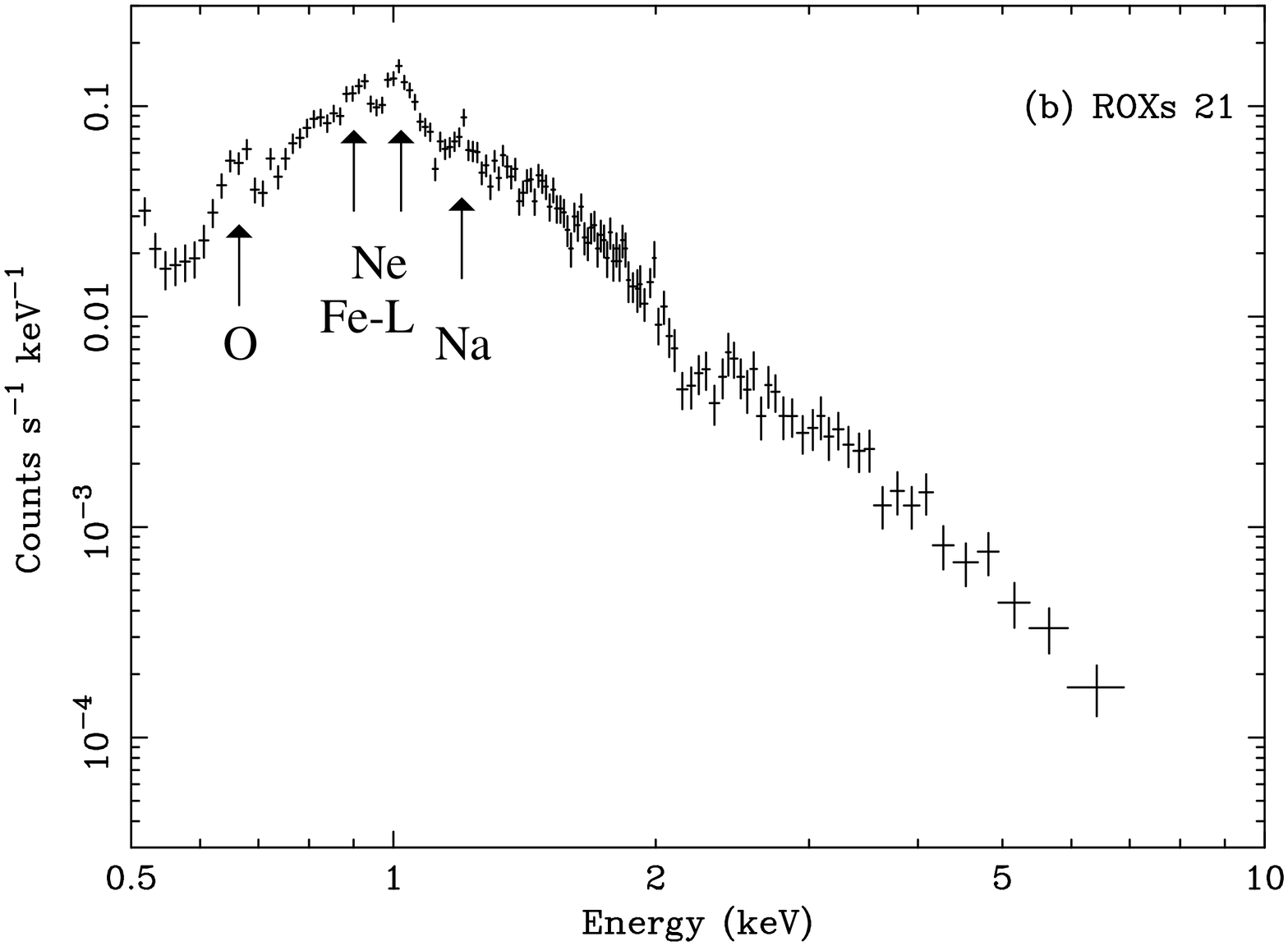}
\plotone{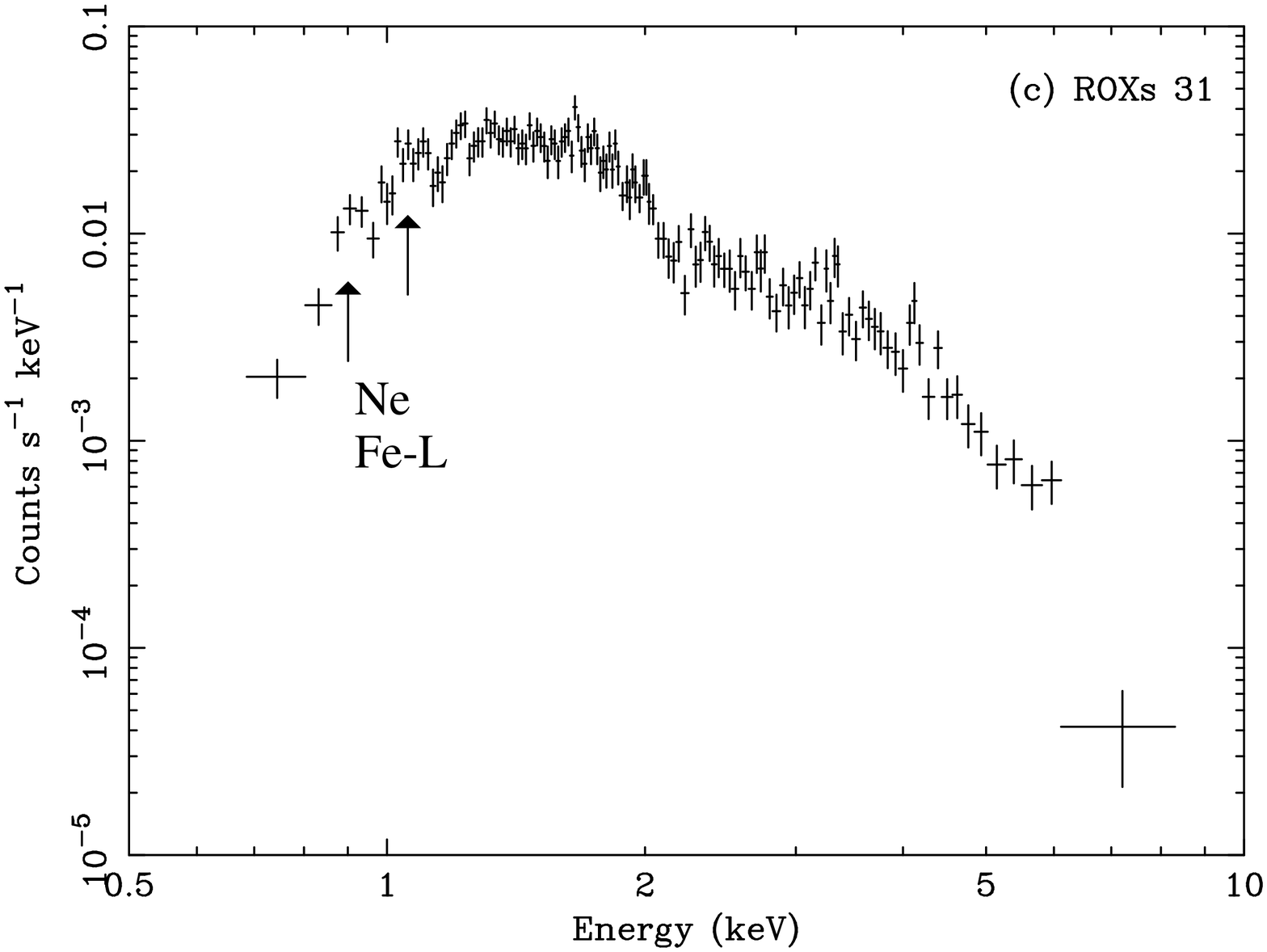}
 \caption[f3a.eps,f3b.eps,f3c.eps]{Time-averaged X-ray spectra of (a)
 DoAr 21, (b) ROXs 21, and (c) ROXs 31 obtained with {\it Chandra}
 ACIS-I (obs C1). The prominent spectral features are indicated by the
 arrows and the corresponding elements. \label{fig:spec}}
\end{figure}

\begin{figure}
 \figurenum{4}
 \epsscale{0.32}
\plotone{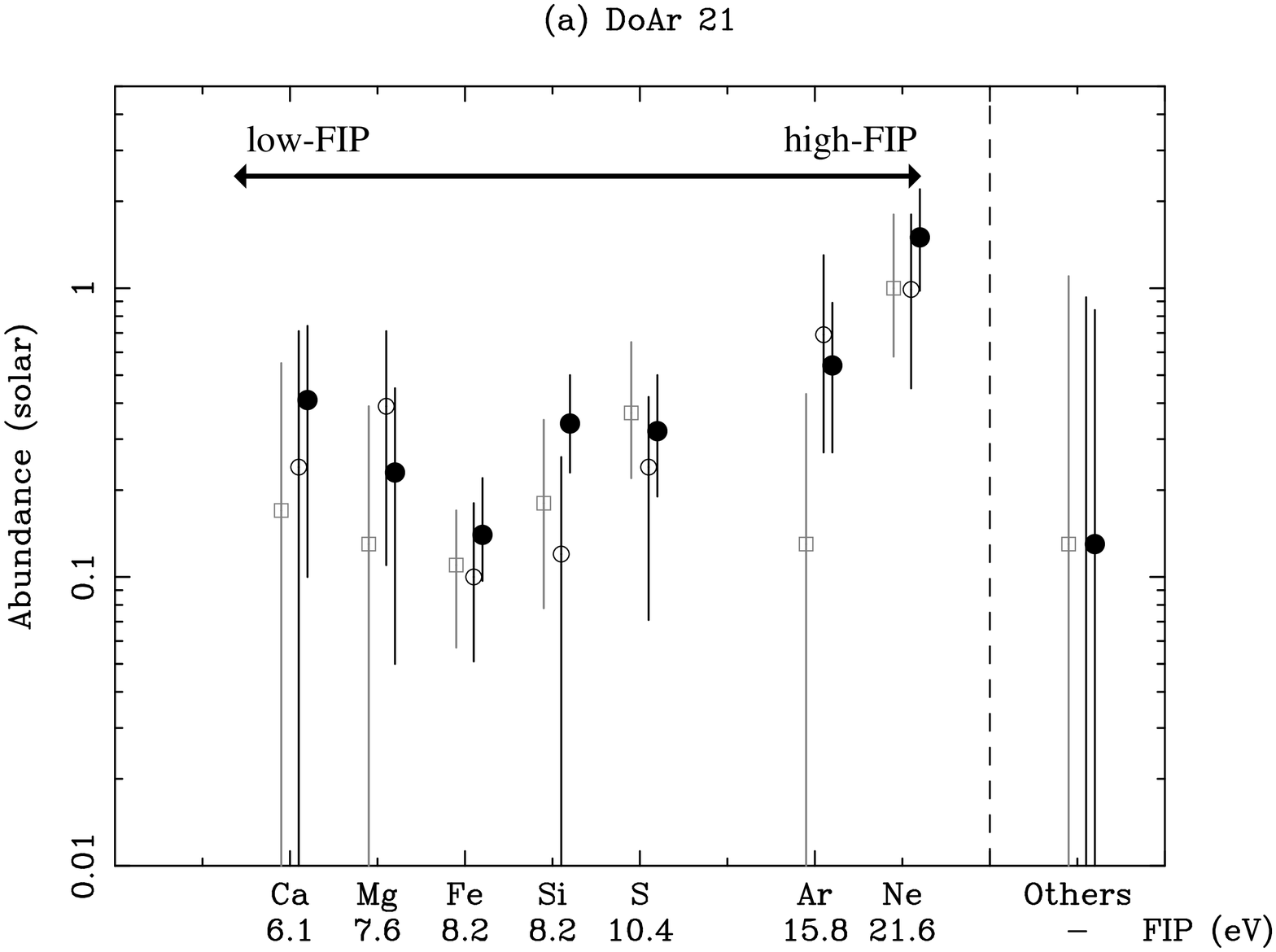}
\plotone{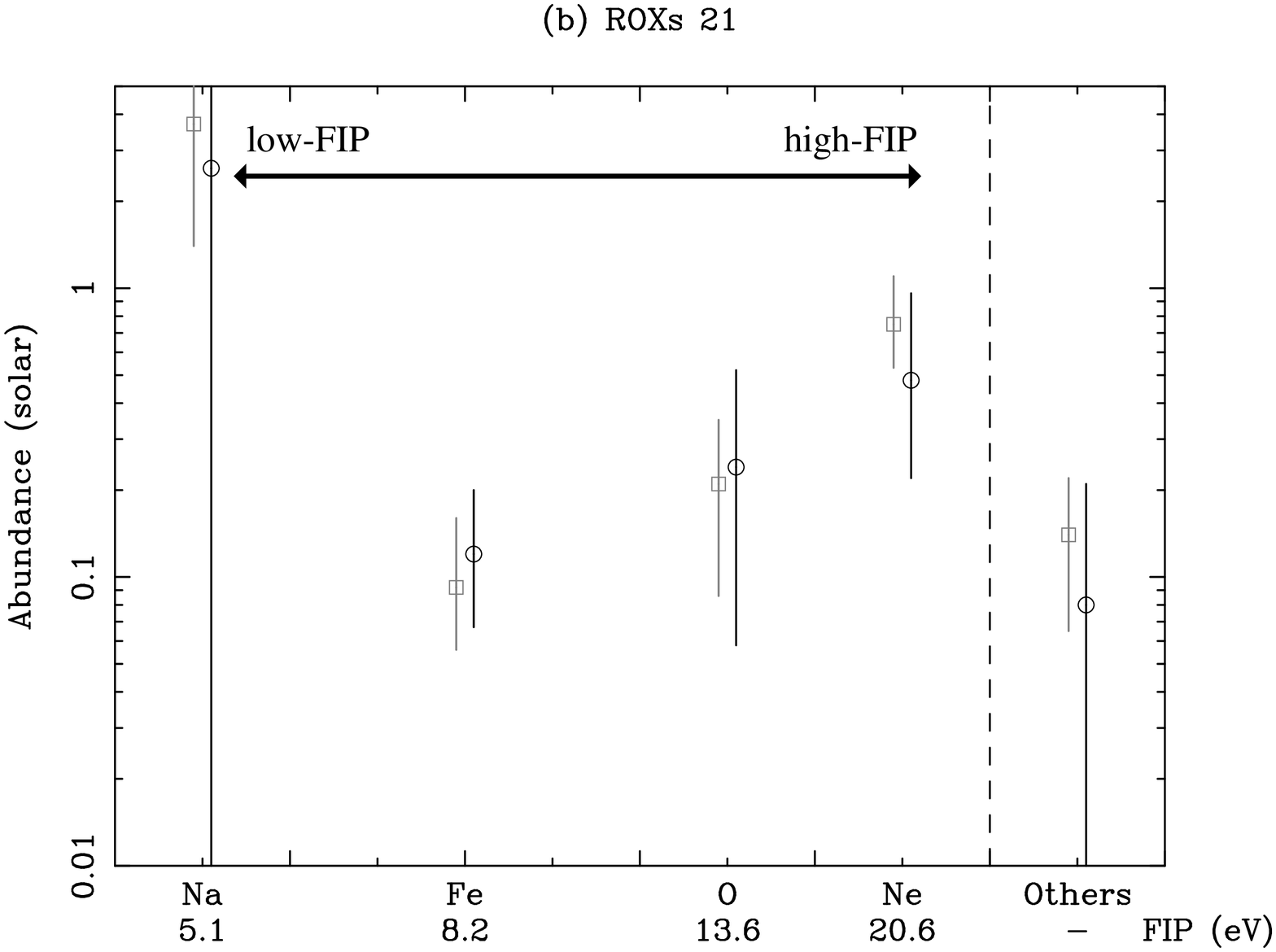}
\plotone{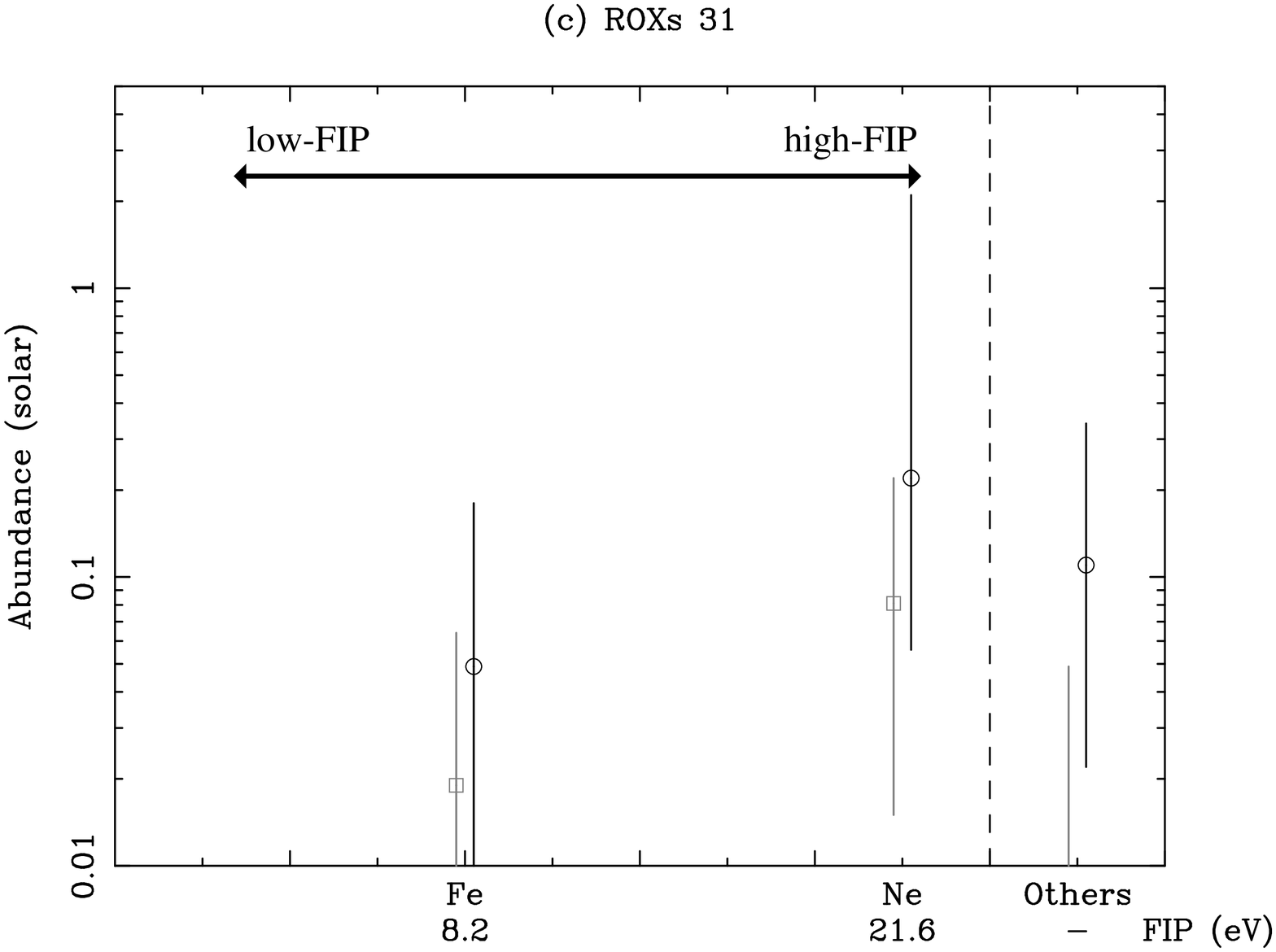}
 \caption[f4a.eps,f4b.eps,f4c.eps]{Elemental abundances of (a) DoAr 21,
 (b) ROXs 21, and (c) ROXs 31. Errors indicate 90\% confidence
 limits. The horizontal axis shows corresponding elements with their FIP
 values (eV). Gray squares, black-open, and black-filled circles
 represent abundances in the quiescent, first and second flare phases,
 respectively. \label{fig:abund_doar21}}
\end{figure}

The background-subtracted ACIS spectra of each source using all the data
of obs C1 are shown in Figure \ref{fig:spec}. Several emission lines
from highly ionized elements such as O, Ne, Na, Mg, Si, S, Ar, Ca, and
Fe are seen, hence the spectra are characteristic of a thin thermal
plasma. We therefore fit the spectra with a thin thermal plasma model
allowing the abundance of each element with prominent lines (indicated
by arrows in Figure \ref{fig:spec}) to be free.  For the study of time
variation of the spectra, we separately fit the flare and quiescent
spectra as shown in Figure \ref{fig:lc_c1}. A single-temperature (1-T)
thin-thermal plasma model (Mewe \etal 1985) is not acceptable for all
the spectra except DoAr 21 and the flare phase of ROXs 31. We therefore
fit two-temperature (2-T) plasma models for ROXs 21 and ROXs 31 and find
acceptable fits. Table \ref{tab:spec} shows the best-fit parameters. All
the elemental abundances are below the solar photospheric values
\citep{Anders1989} except for Ne in DoAr 21 and Na in ROXs 21. The
best-fit abundances are displayed in Figure \ref{fig:abund_doar21} with
the order of the FIP.

\begin{figure}
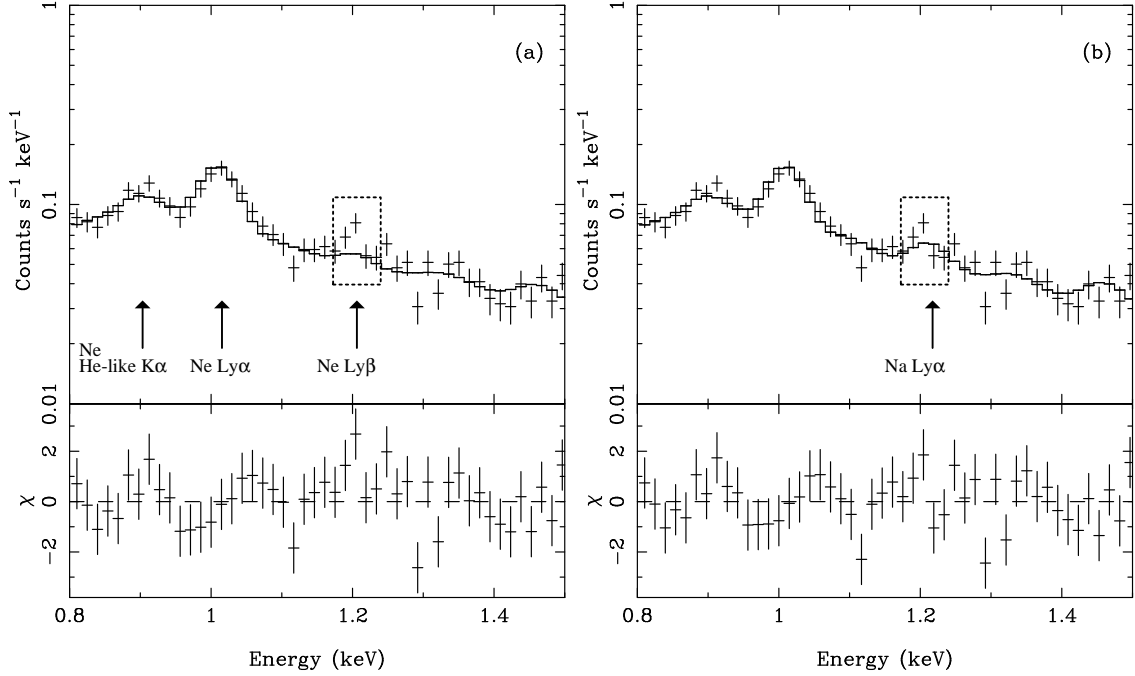

 \figurenum{5}
 \epsscale{0.45}
\plotone{f5a.eps}
\plotone{f5b.eps}
 \caption[f5a.eps,f5b.eps]{Close-up view of the fitting results of ROXs
 21 in phase Q, in which the Na abundance (a) is varied correctively
 with the ``other'' elements and (b) is treated as a free parameter. The
 positions of K-shell lines from (a) Ne and (b) Na are shown by
 arrows. The prominent feature near the Lyman-$\alpha$ line of Na is
 also indicated by dashed squares.\label{fig:spec_roxs21_1keV}}
\end{figure}

We investigate the Na overabundance seen in ROXs 21 in further detail,
because Na is a relatively rare element compared with the major 4N
nuclei and the Na Lyman-$\alpha$ line lies near the Lyman-$\beta$ line
of the more abundant element Ne. Figure \ref{fig:spec_roxs21_1keV}a
shows the best-fit result when the Na abundance is correctively varied
with the ``other'' elements. A significant data excess is found at
$\approx$1.2 keV, near the Lyman-$\beta$ line of Ne (1.211 keV). The
intensity ratio of Lyman-$\beta$ to -$\alpha$ lines of Ne should
increase with increasing plasma temperature, but stay almost constant at
$\approx$ 7 for plasma temperatures higher than 1 keV (Mewe \etal 1985).
Therefore 1/7 of the Lyman-$\alpha$ flux should be an upper-limit to the
Lyman-$\beta$ line flux.  Since the flux of the Ne Lyman-$\alpha$ line
is fairly strong, the Lyman-$\beta$ line flux is well constrained as
seen in Fig 5a. Thus the excess flux should originate from another
element; the Lyman-$\alpha$ line of Na (1.236 keV) is the best
candidate. Allowing the Na abundance to be a free parameter improved the
fit as demonstrated in Figure \ref{fig:spec_roxs21_1keV}b. The reduction
of $\chi^2$ ($d.o.f.$) from 114 (97) to 108 (96) is significant with
$\sim$98 \% confidence level for the $F$-test \citep{Bevington1992}.

\begin{figure}
 \figurenum{6}
 \epsscale{0.45}
\plotone{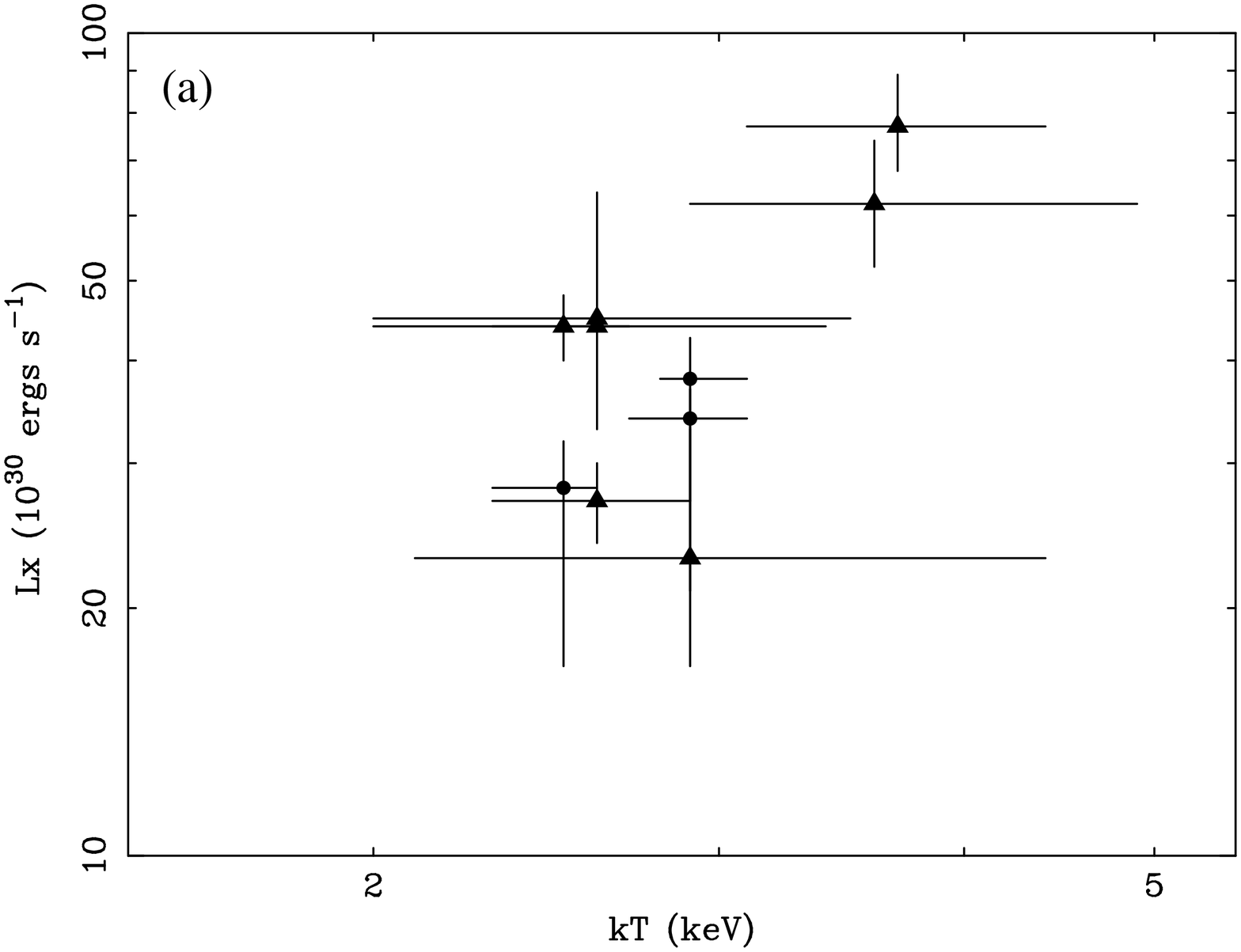}
\plotone{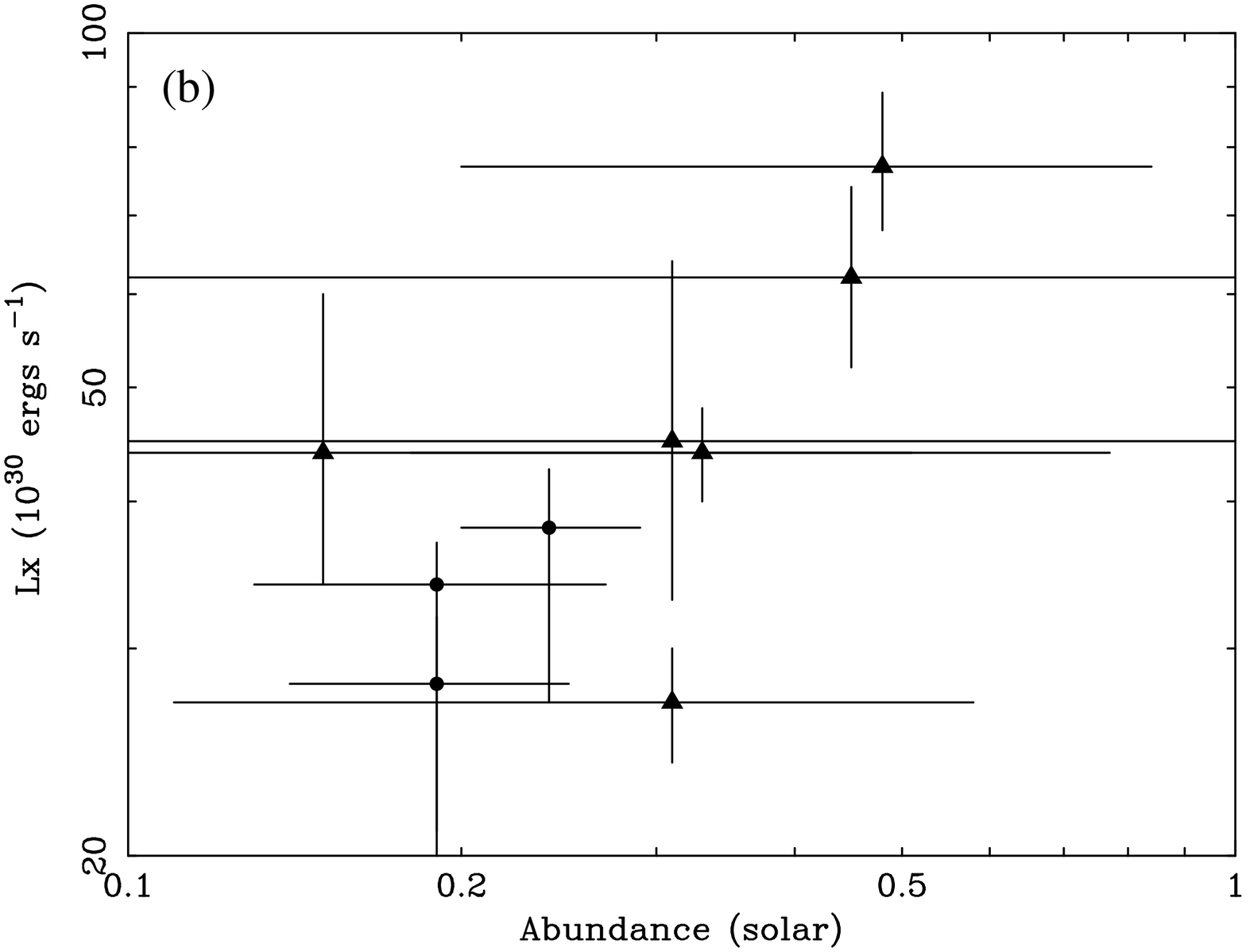}
 \caption[f6a.eps,f6b.eps]{Plot of the X-ray luminosity and the plasma
 temperature (a) and the mean abundance (b) of DoAr 21.  Circles and
 triangles represent the data of ACIS (obs C1) and of GIS (obs
 A1--A3). Errors indicate the 90\% confidence
 limits. \label{lx_kt_abund}}
\end{figure}

Using the GIS spectra of DoAr 21, we investigate the long-term
($\sim$yr) behavior of the spectral parameters. Since the energy
resolution does not allow us to separate the emission lines from the key
elements, we correctively varied the abundances. The best-fit luminosity
shows long-term variability in the range $ (2-8)\times10^{31}$ erg
s$^{-1}$.  The best-fit temperatures and abundances also vary from 2.5
to 4 keV and 0.15 to 0.5 solar, respectively; both are possibly
correlated to the X-ray luminosity as are shown in Figure
\ref{lx_kt_abund}a and \ref{lx_kt_abund}b.  The $N_{\rm H}$ value is
consistent with being constant, although weak variability within a
factor $\approx$ 2 cannot be rejected.  We note that although the 1-T
model in the medium resolution X-ray spectra may lead to an error in
abundance by a factor $\sim$2 \citep {Reale2001}, such an effect is
significantly reduced with 2-T or multi-T models. Since we used 2-T
models for 2 out of the 3 sources, the artificial abundance uncertainty
should be better constrained.

\subsection{Giant Flare from ROXs 31}

\begin{figure}
 \figurenum{7}
 \epsscale{0.45}
\plotone{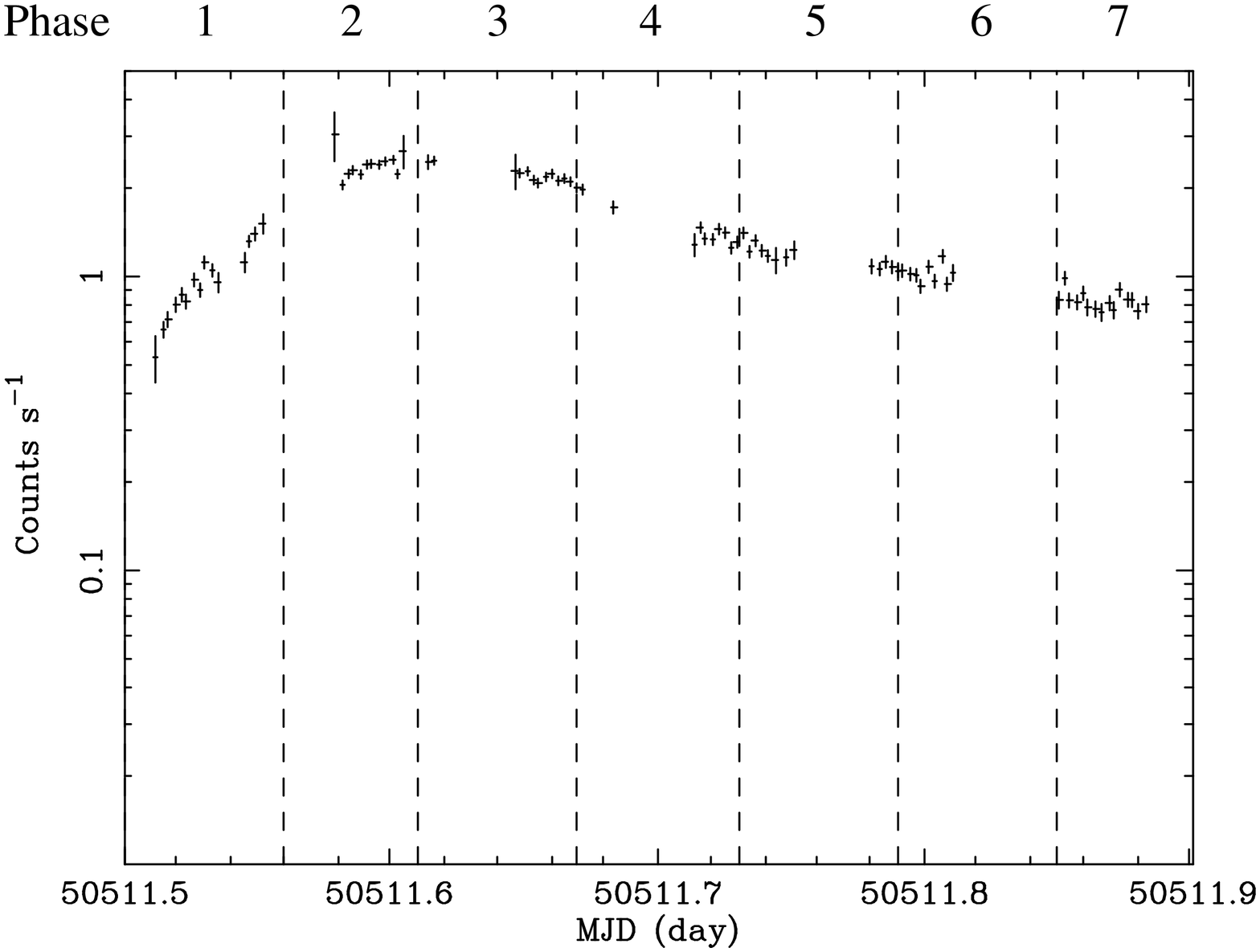}
\plotone{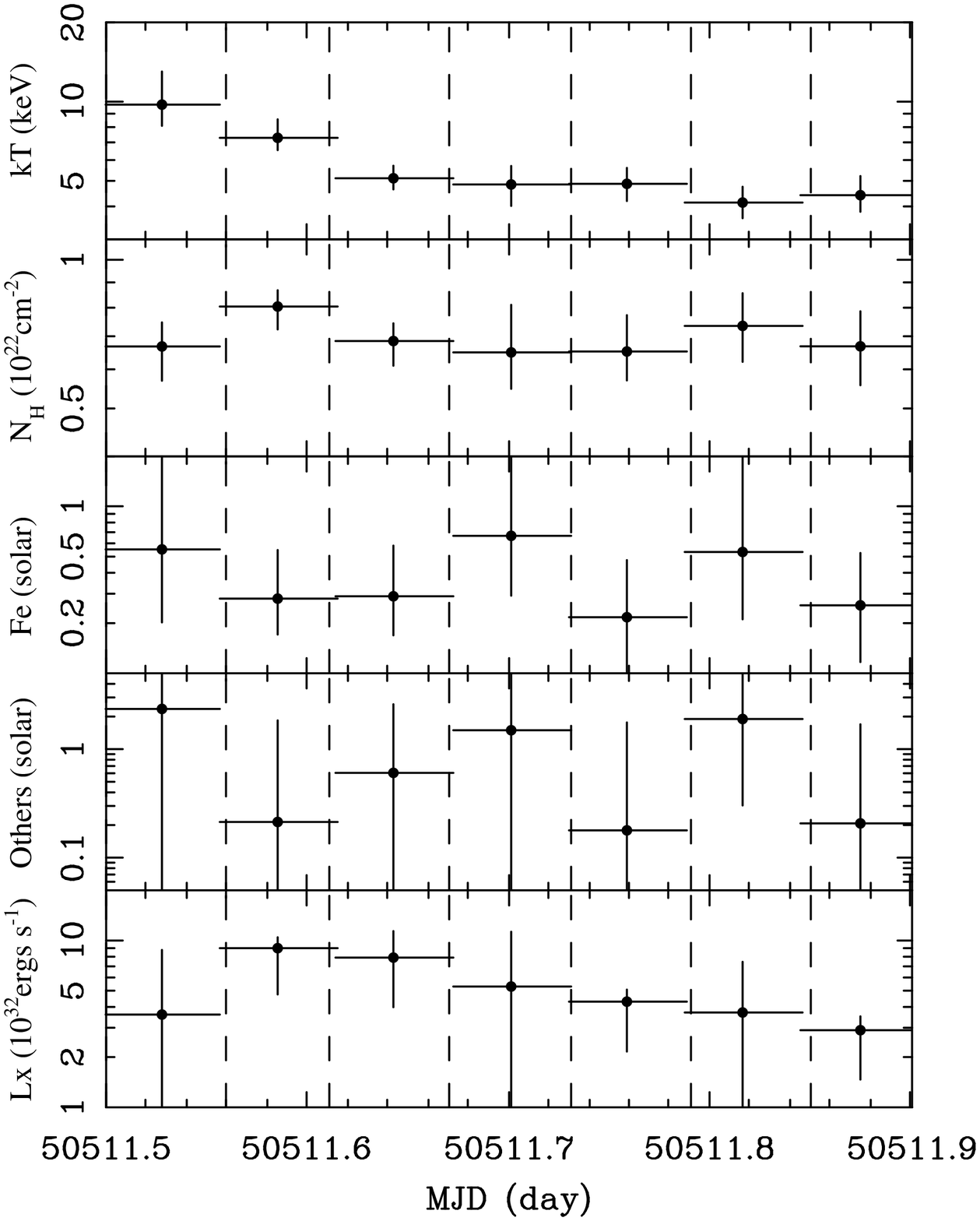}
 \caption[f7a.eps,f7b.eps]{(left)~A magnified view of the light curve of
 ROXs 31 during the giant flare in 0.5--9.0 keV (Figure
 \ref{fig:lc_a1-3}c). (right)~Time profiles of the best-fit temperature,
 absorption column, abundances of Fe and other elements, and X-ray
 luminosity of ROXs 31 during the giant flare in obs A2. Errors indicate
 the 90\% confidence limits. \label{fig:flare_roxs31}}
\end{figure}

For the study of the spectral evolution during the flare, we make and
fit time-sliced GIS spectra of the giant flare from ROXs 31 in obs A2
(Figure \ref{fig:lc_a1-3}c).  The time intervals are shown in Figure
\ref{fig:flare_roxs31} (left). The spectra show strong K$\alpha$
emission of He-like Fe at $\approx$6.7 keV, hence we let the Fe
abundance vary.  The flare spectra are well fitted with a 1-T model. The
best-fit parameters for each time interval are shown in Figure
\ref{fig:flare_roxs31} (right). As shown in the figure, this flare has
an unusual time profile. The light curve increases slowly over $\sim$15
ks (see also Skinner 2000). The temperature reaches its peak value
before the flux maximum, and stays nearly constant during the decay
phase.

The absorption ($N_{\rm H}$) and abundances dramatically change from
quiescent to flare; $N_{\rm H}$ in the flare is
$\sim$7.1$\times$10$^{21}$ cm$^{-2}$, $\sim$2 times smaller than that of
the quiescent phase (obs C1) of 1.7$\times$10$^{22}$ cm$^{-2}$, and the
abundances increase by a factor of 3--30 from quiescent (Z = 0.01--0.1
Z$_{\odot}$) to the giant flare (0.26--0.29). These parameters, however,
do not change during the flare.

\section{DISCUSSION}

\subsection{Flare Rate}

We have detected six (DoAr 21), three (ROXs 21) and two (ROXs 31) flares
under our flare criterion (see \S3.2).  The total exposure time of the
four observations is $\sim$7 days, hence the flare rate is one per 1.2,
2.3, and 3.5 days for DoAr 21, ROXs 21 and ROXs 31, respectively.  We
also confirmed the high flare rate of DoAr 21 in obs C2, although the
data suffer from photon pile-up. Two possibilities could account for the
higher rate in DoAr 21: differences of the energy band (DoAr 21:
0.5--9.0 keV, versus ROXs 21 and ROXs 31: 0.5--1.5 keV in obs A1--A3)
and the mean count rate (DoAr 21: $\sim$0.4 counts s$^{-1}$, versus ROXs
21 and ROXs 31: 0.04--0.09 counts s$^{-1}$ in obs C1). In fact, limiting
the GIS data of DoAr 21 to 0.5--1.5 keV reduces the number of flares
(under our criterion) to two (F2 in obs A2 and F in obs A3), similar in
number to those of ROXs 21 and ROXs 31.  Also, the higher count rate of
DoAr 21 results in a higher sensitivity to smaller amplitude flares that
can be detected under our criterion (\S3.2). Indeed, both flares in
Figure \ref{fig:lc_c1}a have smaller amplitudes than those in Figure
\ref{fig:lc_c1}b and c.  Conversely, at similar sensitivities, the flare
rate of ROXs 21 and ROXs 31 may be comparable to that of DoAr 21, in
spite of different ages (DoAr 21, $\sim$10$^5$ yr; ROXs 21 and ROXs 31,
$\sim$10$^6$ yr) and different structure (ROXs 21 and ROXs 31 are
binaries, while DoAr 21 may be a single star).

The typical flare rate of X-ray sources in the Taurus-Auriga-Perseus
region (Stelzer \etal 2000) is 1/(4--5) days (assuming a typical decay
time scale of 1 hour). Therefore we predict significantly higher flare
rates than that reported previously.  The higher duty ratio may be
primarily due to the extended sensitivity in the hard X-ray band ($>$1.5
keV) and their brightness, because the flare activity (flux increase) is
clearer in the harder X-ray band and/or for brighter sources as we have
already demonstrated for DoAr 21 in the previous paragraph and in Figure
\ref{lx_kt_abund}a.

\subsection{Interpretation of the Temperature Structure}

The high-quality spectra of the long {\it Chandra} ACIS exposure reveal
that the 2-T models gives a better fit for overall spectra than the 1-T
models for ROXs 21 and ROXs 31. This supports previous 2-T model fits
for some fraction of other bright TTSs \citep{Carkner1996,
Preibisch1997, Costa2000, Ozawa2000, Tsujimoto2001}.  DoAr 21, on the
other hand, displays a simple 1-T spectrum with $kT \sim$3 keV; hence it
has no additional soft component.  Since the age of DoAr 21
($\sim$10$^5$ yr) is younger than that of ROXs 21 and ROXs 31
($\sim$10$^6$ yr) \citep {Nurnberger1998}, coupled with the result of
\citet{Tsujimoto2001} that 2-T spectra are found more often in older
TTSs than in younger protostars, we speculate that the soft component,
probably a relatively steady corona, is generated gradually as the
system increases in age, finally reaching solar-like corona. In this
scenario, the hard component would be the sum of unresolved flares.

\subsection{Chemical Composition of the X-ray-emitting Plasma}

The coronal abundances of the TTSs are sub-solar, consistent with the
previous results.  From Figure \ref{fig:abund_doar21}, we see that both
high-FIP (Ne and Ar) and low-FIP (Na, Mg and Ca) elements show higher
abundances than the other elements (the IFIP and FIP effects).

For the abundances in solar corona, the FIP-effect appears in elements
with FIPs below 10 eV; these elements are collisionally ionized in the
photosphere at 6000--7000 K temperature, and would be preferentially
transferred to the upper coronal region by electric fields
\citep{Feldman1992}. DoAr 21 and ROXs 21 are a K0 star and a binary of
K4 with M2.5, respectively, and hence have photospheric temperatures of
4000--5000 K, 0.6--0.8 times that of the solar photosphere
\citep{Nurnberger1998}.  Therefore the FIP-effect energy-limit of 10 eV
should be shifted to 6--8 eV, which is near Mg (FIP = 7.8 eV) and Ca
(6.1 eV), but well above Na (5.1 keV). Hence, the abundance enhancements
of Mg, Ca and Na provide independent evidence supporting the
universality of the FIP-scenario proposed by \citet{Feldman1992}.

The X-ray luminosities of our TTS sample are a few $\times$ $10^{30-31}$
ergs s$^{-1}$, which is significantly higher than typical FIP sources of
$ \le 10^{29}$ ergs s$^{-1}$ (such as the sun, $\pi^1$ UMa, and $\chi^1$
Ori; G\"udel \etal 2001c).  The high luminosity may also be responsible
for the IFIP effect we observed. A related issue would be the dramatic
abundance increase in the giant flare of ROXs 31.

A scenario for the IFIP effect was proposed by \citet{Guedel2001c}. A
``downward-pointing'' electric field is produced by the high-energy
non-thermal electrons during X-ray active phase, and suppresses the
transfer of the ionized low-FIP elements into the corona. The electrons
also loose their kinetic energy at the photosphere by the collisional
process and then heat up the cool elements, making the abundance
increase during the flare. The radio gyro-synchrotron emission from DoAr
21 and ROXs 31 further supports this idea, because this is the direct
evidence for the presence of the high-energy electrons. Since the
abundance increase should be correlated with the non-thermal radio flux
variations, simultaneous radio and X-ray observations would help to
constrain physical models.

Another possibility is photo-ionization by coronal X-rays. This was
proposed to explain a solar flare in which Ne is enhanced relative to O
\citep {Shemi1991,Schmelz1993}. The IFIP effect is seen in the X-ray
brightest stars with $L_{X}$ = 10$^{29}$--10$^{32}$ ergs s$^{-1}$ (HR
1099, Brinkman \etal 2001; AB Dor, G\"udel \etal 2001a; YY Gem, G\"udel
\etal 2001b; II Pegasi, Huenemoerder \etal 2001; and TW Hydrae, Kastner
\etal 2001), in which large X-ray photo-ionization may occur.  The
abundances increase in the X-ray flare may also be explained by this
scenario.

\subsection{Interpretation of the Giant Flare of ROXs 31}

The unusual giant flare from ROXs 31 shows 1-T plasma emission, in
contrast to the quiescent phase. The flux of the soft component in the
quiescent phase is 10--100 times lower than that of the flare flux.
Hence it is highly possible that the soft component does exist during
the flare but is hidden by the more luminous harder component.

The peak luminosity of $\sim$10$^{33}$ ergs s$^{-1}$ and temperature of
$\sim$10 keV are comparable to the giant flare of another TTS binary
system V773 Tau \citep{Tsuboi1998}. However, the time profile of the
plasma temperature is different from that of V773 Tau; the temperature
arrives at the peak value preceding the flare peak, and, in the decay
phase, stays nearly constant at 4--5 keV. These imply the existence of
successive energy release due to the magnetic reconnection and that the
generation and heating of the plasma occur even in the decay phase, as
suggested by \citet{Tsuboi1998}; \citet{Schmitt1999};
\citet{Favata1999}. Such a phenomenon is also seen in long duration
event (LDE) flares of the Sun \citep{Isobe2001}.

The abundance and absorption during the giant flare are systematically
larger and smaller than those in the quiescent phase (Table
\ref{tab:spec}). This could be due to evaporation of photosphere and
environments by high-energy electrons, which may be produced just after
magnetic reconnection.  Evaporation by the flare X-rays may not be
important, because the abundances and absorption do not change much
during the flare. Thus, in the case of the IFIP effect in the giant
flare of ROXs 31, the high-energy electron origin is more likely than
X-ray irradiation scenario.

\section{SUMMARY}

We analyzed the {\it Chandra} and {\it ASCA} data of the three
X-ray-bright TTSs, DoAr 21, ROXs 21, and ROXs 31. Based on the timing
and spectral features, we derived several results as follows;

\begin{enumerate}

 \item Six flares in the 0.5--9 keV band are detected from DoAr 21
       during the total duration of $\sim$7 days. Also, three and two
       flares are detected from ROXs 21 and ROXs 31, respectively, with
       {\it Chandra} in 0.5--9.0 keV and {\it ASCA} in 0.5--1.5 keV.
       Converting to the 0.5-9 keV band sensitivity, the flare numbers
       become nearly the same as DoAr 21. Thus the flare rate of these
       TTSs may be extremely high (about one per one day).
 \item The spectra of DoAr 21 are well fitted with a 1-T model, while
       those of ROXs 21 and ROXs 31 need additional soft components.
       This can be interpreted as a gradual generation of the soft
       component with the increasing age.
 \item We found relative abundance enhancements in both the low-FIP (Na,
       Mg and Ca) and high-FIP (Ne and Ar) elements.
 \item An unusual giant flare is detected from ROXs 31. The temperature
       remains high (4--5 keV) even during the decay phase, suggesting a
       successive energy release.  A large abundance and absorption jump
       from quiescent to flare is found, which may suggest a large
       contribution to gas evaporation from the photosphere and
       environments by non-thermal electrons.
\end{enumerate}

\acknowledgements{The authors express their thanks to Dr. Thomas
Preibisch for critical refereeing and useful comments. The authors also
acknowledge Dr. Franz Bauner, Dr. Yohko Tsuboi, Hiroaki Isobe, and Taro
Morimoto for useful discussions and comments. The {\it Chandra} data
were obtained through the Chandra X-ray Observatory Science Center (CXC)
operated for NASA by the Smithsonian Astrophysical Observatory. The {\it
ASCA} data were obtained through the High Energy Astrophysics Science
Archive Research Center Online Service provided by the NASA/Goddard
Space Flight Center. K.I.\ and M.T.\ are financially supported by JSPS
Research Fellowship for Young Scientists.}

%\clearpage

\onecolumn
%\clearpage

\clearpage

\begin{deluxetable}{llllc}
 \tabletypesize{\scriptsize}
 \tablewidth{0pt}
 \tablenum{1}
 \tablecaption{Observation Log \label{tab:obs}}
 \tablehead{\colhead{Obs.ID} & \colhead{Detector} & \colhead{Sequence
 ID} & \colhead{Date} & \colhead{Exposure} \\
 & & & & \colhead{(ks)}}
 \startdata
 %%%%%%%%%%%%%%%%%%%%
 A1 & {\it ASCA}/GIS & 20015010 & 1993 Aug 20--20 & \phn37.7 \\
 A2 & {\it ASCA}/GIS & 25020000 & 1997 Mar 2--4   & \phn93.1 \\
 A3 & {\it ASCA}/GIS & 96003000 & 1998 Aug 13--15 & \phn74.7 \\
 C1 & {\it Chandra}/ACIS & 200060 & 2000 Apr 13--14 & 100.6 \\
 %%%%%%%%%%%%%%%%%%%%
 \enddata
\end{deluxetable}

\begin{deluxetable}{lllllllll}
 \tabletypesize{\scriptsize}
 \rotate
 \tablewidth{0pt}
 \tablenum{2}
 \tablecaption{Best-Fit Parameters in obs C1\tablenotemark{a}
 \label{tab:spec}}
 \tablehead{\colhead{Source} & \multicolumn{3}{c}{DoAr 21} &
 \multicolumn{2}{c}{ROXs 21} & \multicolumn{2}{c}{ROXs 31} \\
 \colhead{Phase\tablenotemark{b}} & \colhead{Q\tablenotemark{c}} &
 \colhead{F1} & \colhead{F2\tablenotemark{c}} & \colhead{Q} &
 \colhead{F} & \colhead{Q} & \colhead{F}}
 \startdata
 %%%%%%%%%%%%%%%%%%%%
 $kT_1$ (keV) & 2.5(2.4--2.6) & 2.9(2.7--3.1) & 2.9(2.8--3.1) &
 0.69(0.65--0.74) & 0.86(0.78--4.5) & 0.23(0.15--0.31) & 0.27(0.21--0.41)
 \\
 $kT_2$ (keV) & \nodata & \nodata & \nodata &
 1.8(1.6--2.1) & 2.6(2.3--3.1) & 1.5(1.2--1.7) & 2.0(1.8--2.5) \\
 $N_{\rm H}$ (10$^{22}$ cm$^{-2}$) & 1.0(0.97--1.1) & 1.0(0.96--1.1) &
 1.10(1.06--1.13) & 0.13(0.079--0.16) & 0.081(0.038--0.11) &
 1.7(1.1--2.1) & 1.7(1.0--2.5) \\
 \\
 Abundances (solar) \\
 \phn O & \nodata\tablenotemark{d} & \nodata\tablenotemark{d} &
 \nodata\tablenotemark{d} & 0.21(0.086--0.37) & 0.24(0.058--0.52) &
 \nodata\tablenotemark{d} & \nodata\tablenotemark{d} \\
 \phn Ne & 1.0(0.58--1.8) & 0.99(0.45--1.8) & 1.5(0.98--2.2) &
 0.75(0.53--1.1) & 0.48(0.22--0.96) & 0.081(0.015--0.22) &
 0.22(0.056--2.1) \\
 \phn Na & \nodata\tablenotemark{d} & \nodata\tablenotemark{d} &
 \nodata\tablenotemark{d} & 3.7(1.4--8.2) & 2.6($<$7.6) &
 \nodata\tablenotemark{d} & \nodata\tablenotemark{d} \\
 \phn Mg & 0.13($<$0.39) & 0.39(0.11--0.71) & 0.23(0.050--0.45) &
 \nodata\tablenotemark{d} & \nodata\tablenotemark{d} &
 \nodata\tablenotemark{d} & \nodata\tablenotemark{d} \\
 \phn Si & 0.18(0.078--0.35) & 0.12($<$0.26) & 0.34(0.23--0.50) &
 \nodata\tablenotemark{d} & \nodata\tablenotemark{d} &
 \nodata\tablenotemark{d} & \nodata\tablenotemark{d} \\
 \phn S & 0.37(0.22--0.65) & 0.24(0.071--0.42) & 0.32(0.19--0.50) &
 \nodata\tablenotemark{d} & \nodata\tablenotemark{d} &
 \nodata\tablenotemark{d} & \nodata\tablenotemark{d} \\
 \phn Ar & 0.13($<$0.43) & 0.69(0.27--1.3) & 0.54(0.27--0.88) &
 \nodata\tablenotemark{d} & \nodata\tablenotemark{d} &
 \nodata\tablenotemark{d} & \nodata\tablenotemark{d} \\
 \phn Ca & 0.17($<$0.55) & 0.24($<$0.73) & 0.41(0.10--0.74) &
 \nodata\tablenotemark{d} & \nodata\tablenotemark{d} &
 \nodata\tablenotemark{d} & \nodata\tablenotemark{d} \\
 \phn Fe & 0.11(0.057--0.17) & 0.10(0.051--0.18) & 0.14(0.097--0.22) &
 0.092(0.056--0.16) & 0.12(0.067--0.20) & 0.019($<$0.064) & 0.049($<$0.18) 
 \\
 \phn Others & 0.13($<$1.1) & 0($<$0.93) & 0.13($<$0.84) &
 0.14(0.065--0.22) & 0.080($<$0.21) & 0.0080($<$0.049) &
 0.11(0.022--0.34) \\
 \\
 $L_X$ (10$^{30}$ ergs s$^{-1}$)\tablenotemark{e} & 28 & 34 & 38 & 1.5 &
 1.9 & 22 & 30 \\
 reduced-$\chi^2$($d.o.f.$) & 0.953(175) & 0.910(150) & 1.03(275) &
 1.12(96) & 1.07(88) & 1.11(67) & 0.931(85) \\
 %%%%%%%%%%%%%%%%%%%%
 \enddata
 \tablenotetext{a}{The 1-T model for DoAr 21 and the 2-T models for ROXs
 21 and ROXs 31 are used. Parentheses indicate the 90\% confidence
 limits.}
 \tablenotetext{b}{See Figure \ref{fig:lc_c1}.}
 %
 %KI: add a footnote
 %
 \tablenotetext{c}{The boundary between phases Q and F2 is not clear
 because of the existence of a rising part (Figure \ref{fig:lc_c1}),
 however there are no significant differences in parameters when the
 boundary is changed.}
 \tablenotetext{d}{These are assumed to be the same as ``Other''
 elements.}
 \tablenotetext{e}{Absorption-corrected luminosity in 0.5--9.0 keV.}
\end{deluxetable}

\end{document}